\documentclass[aps,prl,twocolumn,showpacs]{revtex4}
\usepackage{graphics,graphicx}
\usepackage{color}
\usepackage{wrapfig}
\usepackage{enumerate}
\usepackage{amssymb,amsmath}
\usepackage{bm}

\usepackage[normalem]{ulem}

\begin{document}
\unitlength = 1mm
\title{
  Numerical evidence of quantum melting of spin ice: quantum-classical crossover
}

\author{Yasuyuki~Kato$^{1,2}$ and Shigeki~Onoda$^{1,3}$}

\affiliation{
  $^1$RIKEN Center for Emergent Matter Science (CEMS), Wako, Saitama 351-0198, Japan\\
  $^2$Department of Applied Physics, University of Tokyo, Bunkyo, Tokyo 113-8656, Japan\\
  $^3$Condensed Matter Theory Laboratory, RIKEN, Wako, Saitama 351-0198, Japan
}

\date{\today}
%
\begin{abstract}
  Unbiased quantum Monte-Carlo simulations are performed on the nearest-neighbor spin-$\frac{1}{2}$ pyrochlore XXZ model with an antiferromagnetic longitudinal and a weak ferromagnetic transverse exchange couplings, $J$ and $J_\perp$. The specific heat exhibits a broad peak at $T_{\mathrm{CSI}}\sim0.2J$ associated with a crossover to a classical Coulomb liquid regime showing a suppressed spin-ice monopole density, a broadened pinch-point singularity, and the Pauling entropy for $|J_\perp|\ll J$, as in classical spin ice. On further cooling, the entropy restarts decaying for $J_\perp>J_{\perp c}\sim-0.104J$, producing another broad specific heat peak for a crossover to a bosonic quantum Coulomb liquid, where the spin correlation contains both photon and quantum spin-ice monopole contributions. With negatively increasing $J_\perp$ across $J_{\perp c}$, a first-order thermal phase transition occurs from the quantum Coulomb liquid to an XY ferromagnet. Relevance to magnetic rare-earth pyrochlore oxides is discussed.
\end{abstract}
\maketitle

A compact U(1) gauge theory hosts dual electric and magnetic monopoles as well as photons, as emerge in non-Abelian gauge theories for grand unified theories~\cite{thooft:74,polyakov:74}. In condensed matter, it is expected to appear at the ground state of the nearest-neighbor spin-$\frac{1}{2}$ XXZ model on the pyrochlore lattice~\cite{hermele2004,moessner:06,banerjee2008}, given by the Hamiltonian,
\begin{eqnarray}
  {\mathcal H}= \sum_{\langle {\bm r},{\bm r}'\rangle} \left[
  J s^{z}_{\bm r} s^{z}_{{\bm r}'}
  +J_{\bot} \left( s^{x}_{\bm r} s^{x}_{{\bm r}'} + s^{y}_{\bm r} s^{y}_{{\bm r}'}\right)
\right],
\label{eq:model}
\end{eqnarray}
with a spin-$\frac{1}{2}$ operator $\bm{s}_{\bm{r}}=(s^x_{\bm{r}},s^y_{\bm{r}},s^z_{\bm{r}})$ at a pyrochlore lattice site $\bm{r}$, and the nearest-neighbor longitudinal ($z$) and transverse ($xy$) exchange couplings $J (> 0)$ and $J_\perp$.
This model, defined in the $C_2$-invariant local spin frames with their $z$ axes pointing inwards to or outwards from the center of the tetrahedron~\cite{onoda:09,onoda:10,onoda:11}, gives the most simplified case of low-energy effective spin models for magnetic rare-earth pyrochlore oxides, e.g., Pr$_2$Ir$_2$O$_7$~\cite{machida:09}, Pr$_2$Zr$_2$O$_7$~\cite{onoda:09,onoda:10,kimura:13}, Yb$_2$Ti$_2$O$_7$~\cite{onoda:11,ross:11,chang:11}, and Tb$_2$Ti$_2$O$_7$~\cite{Gardner2001,Mirebeau2007,Molavian2007,onoda:10,Taniguchi2013,Curnoe2013}.

The particular limit $J_\perp=0$ of the model is reduced to the nearest-neighbor classical spin ice (CSI) model. It involves a macroscopic degeneracy of the ground states satisfying the 2-in, 2-out spin ice rule~\cite{bernal:33,harris:97} and the Pauling residual entropy $S_{\mathrm{P}}=\frac{1}{2}\ln\frac{3}{2}$~\cite{pauling}, as observed in Ho$_2$Ti$_2$O$_7$~\cite{cornelius:01} and Dy$_2$Ti$_2$O$_7$~\cite{ramirez:99}. This CSI has been well understood in terms of a classical Coulomb (CC) phase physics in a gauge theory on the dual diamond lattice~\cite{isakov:04,henley:05,moessner:06,castelnovo:08}: the Hamiltonian is given by $\frac{J}{2}\sum_{\bm{R}_\sigma}n_{\bm{R}_\sigma}^2$ with the static gauge charge, $n_{\bm{R}_\sigma}=\sigma\sum_{\mu=0,\cdots,3}s^z_{\bm{R}_\sigma+\sigma\bm{b}_\mu/2}$, defined at the center $\bm{R}_\sigma$ of the tetrahedron, where $\sigma=\pm$ and $\bm{b}_\mu$ denote the sublattice index of the diamond lattice and the four nearest-neighbor diamond lattice vectors, respectively. 
On cooling down to zero temperature, the population of this gauge charge, dubbed a spin ice monopole~\cite{castelnovo:08}, vanishes and the spin correlations become of the dipolar form~\cite{isakov:04}.

In the simplest quantum spin ice (QSI) model \eqref{eq:model} with nonzero $J_\perp$, the gauge charge acquires a quantum kinematics as a QSI monopole, which is a spin-$\frac{1}{2}$ bosonic spinon playing a role of scalar Higgs fields in the U(1) gauge theory~\cite{lee2012}. This kinematics completely lifts the degeneracy of the spin ice manifold, leading to quantum melting of spin ice~\cite{onoda:09}. A degenerate perturbation theory about $J_\perp$ yields a bosonic U(1) quantum spin liquid having deconfined dual gauge charges and linearly dispersive gapless ``photons''~\cite{hermele2004}. This prediction was partially tested by quantum Monte-Carlo (QMC) simulations~\cite{banerjee2008} which found a consistency of the spin correlations with those of ``photons'' for small $J_\perp$ at low temperatures, while the bosonic U(1) quantum spin liquid is replaced by an XY ferromagnet (XY-FM) at large negative $J_\perp$. For a pure lattice U(1) gauge model obtained by projecting Eq.~\eqref{eq:model} onto the spin ice manifold, a Green function Monte-Carlo study showed that a scaling of the ground-state energy with the U(1) gauge flux supports the emergence of ``photons''~\cite{shannon2012}.

\begin{figure}[htbp]
  \includegraphics[trim = 0 0 0 0, clip,width=7.5cm]{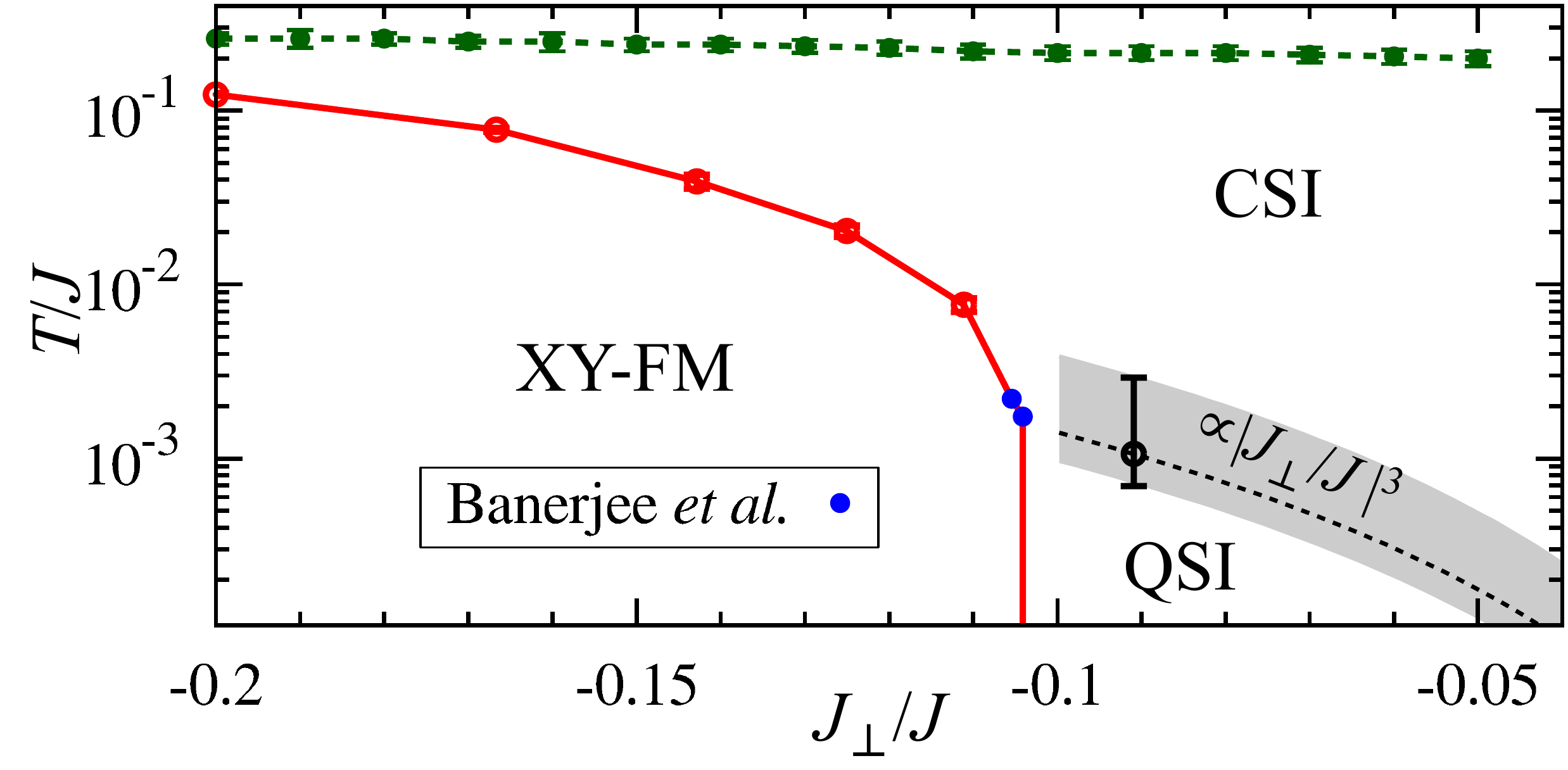} 
  \caption{
    (Color online)
    Finite-temperature phase diagram for $J_{\bot} < 0$, obtained with QMC simulations. Below the phase boundary (solid line) the transverse ($xy$) component of spins are ferromagnetically ordered with finite order parameter $\langle s^+ \rangle$ and spin stiffness $\rho_S$. The blue dots are extracted from Ref.~\cite{banerjee2008}. The dashed lines indicate the crossover temperatures $\frac{T_{\mathrm{CSI}}}{J}$ and $\frac{T_{\mathrm{QSI}}}{J}$ estimated from the position of the broad peaks in the specific heat. The lower-$T$ (black) dashed line interpolates our results at $\frac{J_\perp}{J}=-\frac{1}{11}$ and the $\frac{J_\perp}{J}=0$ limit~\cite{hermele2004}. 
    \label{fig:1}
  }
\end{figure}

On the other hand, finite-temperature properties remain open. A decrease of the entropy below $S_{\mathrm{P}}$ has been observed at very low temperatures in Dy$_2$Ti$_2$O$_7$~\cite{pomaranski:13}, Yb$_2$Ti$_2$O$_7$~\cite{chang:13}, Pr$_2$Zr$_2$O$_7$~\cite{kimura:13}, and Pr$_2$Ir$_2$O$_7$~\cite{tokiwa:14}, whose nature has not been fully understood yet. Some could be ascribed to an onset of either a crossover from a CC liquid to a quantum Coulomb (QC) liquid~\cite{hermele2004} or a transition/proximity to a long-range order. Recent mean-field calculations based on Wilson's idea~\cite{wilson:74}, which violates Elitzur's theorem prohibiting a broken local gauge invariance~\cite{elitzur:75},  in terms of the compact Abelian lattice Higgs model description highlighted a possibly spurious first-order thermal phase transition to a deconfined phase that does not break any physical symmetry but hosts emergent gauge fields~\cite{savary:13}. Now, unbiased calculations for finite-temperature properties of even the simplest QSI model have been called for.


In this Letter, we reveal a finite-temperature phase diagram of the simplest QSI model~\eqref{eq:model}, uncovering two successive crossovers and a single phase transition shown with dashed curves and a solid curve in Fig.~1. On cooling, the system first crosses over to a CC liquid or CSI regime with the entropy $S\sim S_{\mathrm{P}}$. For $J_\bot>J_{\bot c}$ with $\frac{J_{\perp c}}{J}=-0.104$~\cite{banerjee2008}, another crossover occurs to a QC liquid or QSI regime where the entropy decays from $S_{\mathrm{P}}$ and spin correlations evolve continuously towards the formation of ``pyrochlore photons'' at the deconfined QC liquid ground state~\cite{hermele2004}. This rules out a possibility of the first-order thermal confinement-deconfinement transition from CSI to QSI at a temperature scale $T\sim\frac{J_\perp^3}{J^2}$~\cite{savary:13}. For $J_\bot<J_{\bot c}$, there occurs a phase transition, which is of the first order at least for $\frac{J_\perp}{J}\ge-\frac{1}{7}$, to the XY-FM~\cite{banerjee2008}.

All the numerical results presented in this Letter are obtained with unbiased worldline QMC simulations based on the path-integral formulation in the continuous imaginary time~\cite{kawashima2004}. To update worldline configurations, we adopt a directed-loop algorithm~\cite{syljuasen2002} in the $\{s^z_{\bm{r}}\}$ basis, with the modification previously introduced for softcore bosonic systems to reduce the computational cost~\cite{kato2007}. To moderate a freezing problem, we employed a thermal annealing in the simulations. We performed typically $\sim$10000 Monte-Carlo sweeps for each temperature.

\begin{figure}[htbp]
  \includegraphics[trim = 0 0 0 0, clip,width=8.0cm]{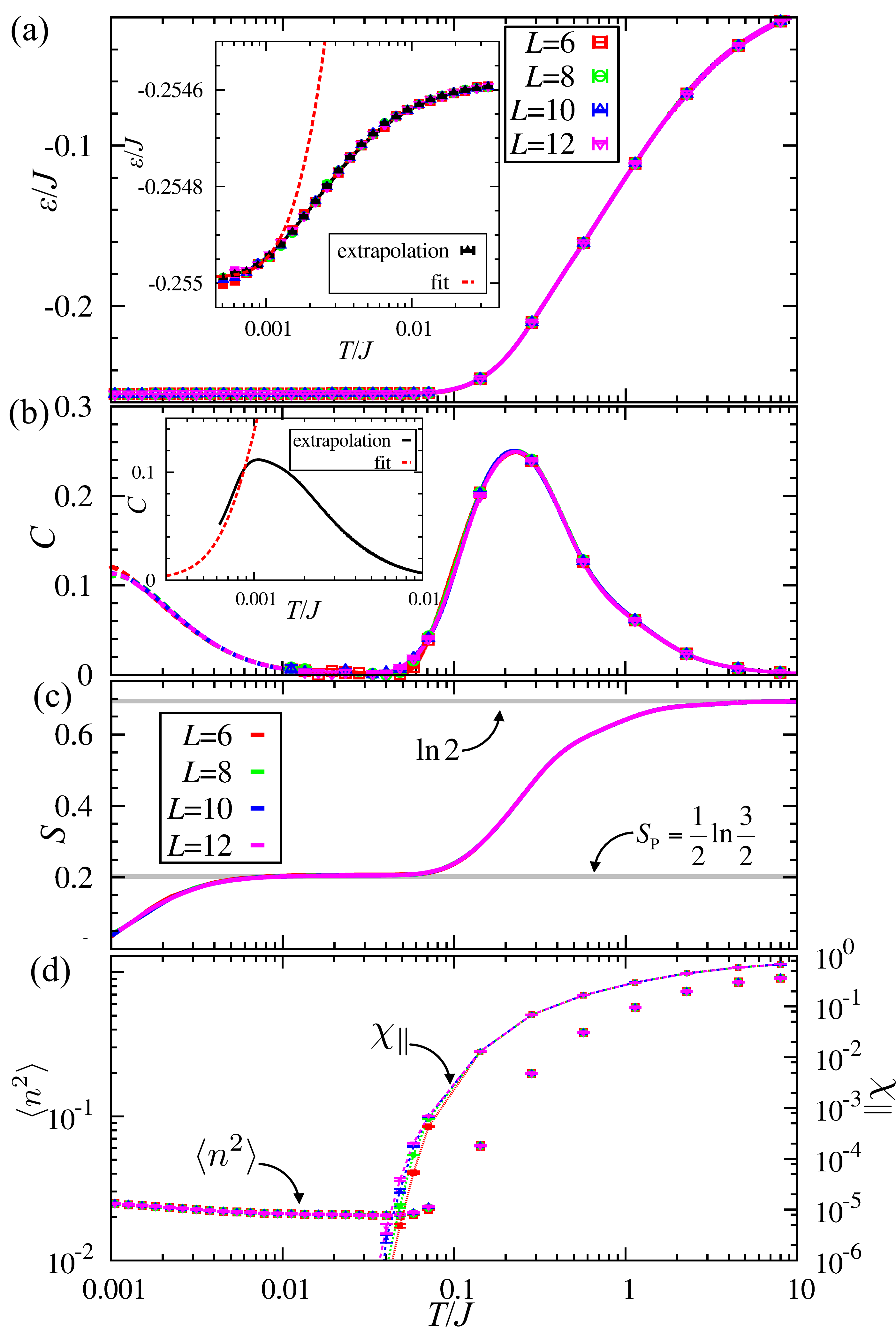} 
  \caption{
    (Color online)
    Temperature dependence of (a) $\varepsilon$, (b) $C$, (c) $S$, and (d) $\langle n^2\rangle$ (left axis) and $\chi_\parallel$ (right axis) for $\frac{J_\perp}{J}=-\frac{1}{11} > (\frac{J_\perp}{J})_c$. Solid and broken curves are the cubic and basis spline interpolations of the QMC data, respectively.
    Insets: black solid and red dashed curves are the basis spline interpolation of the QMC data extrapolated to $L\to\infty$~\cite{supplement} and its fit to the photon contribution $\varepsilon_{\mathrm{photon}}(T)$ to the energy density, respectively, in (a), while their temperature derivatives are shown in (b). For magnified views~\cite{supplement}.
    \label{fig:2}
  }
\end{figure}

\begin{figure*}[htbp]
  \includegraphics[trim = 0 145 0 150, clip,width=0.8\textwidth]{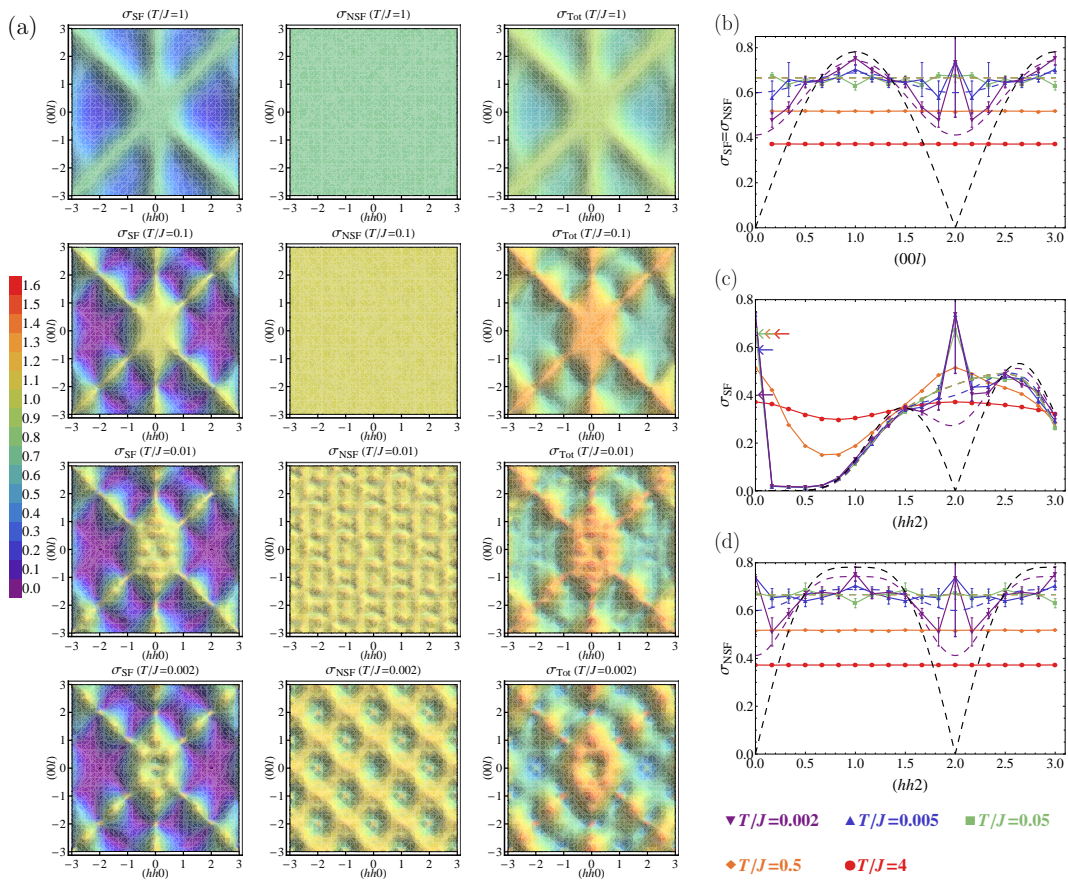}
  \caption{
    (Color online)
    QMC results on $\sigma_{\mathrm{SF}}$, $\sigma_{\mathrm{NSF}}$, and $\sigma_{\mathrm{Tot}}$ for $\frac{J_\perp}{J} = -\frac{1}{11}$ with $L=12$. (a) Profiles on the $\bm{k} = \frac{2\pi}{a} (h,h,l)$ plane at $\frac{T}{J}=4$--$0.002$. (b) $\sigma_{\mathrm{SF}}=\sigma_{\mathrm{NSF}}$ along $\bm{k}=\frac{2\pi}{a}(0,0,l)$ and (c, d) $\sigma_{\mathrm{SF}}$ and $\sigma_{\mathrm{NSF}}$ along $\bm{k}=\frac{2\pi}{a}(h,h,2)$. Dashed curves denote the cases for a noncompact pure U(1) gauge theory on the pyrochlore lattice~\cite{benton2012} with the ``light'' velocity $c=1.49(4)\frac{ag}{\hbar}$, and the black dashed curves are for $T\to0$. The results at $\frac{T}{J}=4$ and $0.5$ merge with those at $\frac{T}{J}=0.05$. The arrows in (c) show the magnitudes of the photon form at $\bm{k}=\lim_{\eta\to0}\frac{2\pi}{a}(0,0,2-\eta)$. 
    \label{fig:3}
  }
\end{figure*}

Let us start with the disordered side $J_\bot > J_{\bot c}$ of the phase diagram. Figure~\ref{fig:2} shows for $\frac{J_\perp}{J}=- \frac{1}{11}$ the temperature dependence of (a) the energy density $\varepsilon \equiv \frac{\langle \mathcal{H} \rangle}{N_s}$ with $N_s = 4 L^3$ being the total number of spins, (b) the specific heat $C \equiv \frac{\partial \varepsilon}{\partial T}$, (c) the entropy $S\equiv \ln 2 - \int_T^{T_{\rm max}} \frac{C}{T} dT$ computed from a numerical integration of the cubic spline interpolation of $\varepsilon$ by taking $\frac{T_{\mathrm{max}}}{J}=20$, and (d) the mean square QSI monopole density $\langle n^2\rangle\equiv\langle n_{\bm{R}_\sigma}^2\rangle$ and the uniform static longitudinal spin susceptibility $\chi_\parallel$, which is nothing but the QSI monopole charge compressibility of an FCC sublattice of the pyrochlore lattice.
A broad specific heat peak appears without significant finite-size effects beyond small statistical errors at $T_{\mathrm{CSI}}\sim 0.2J$. On cooling across $T_{\mathrm{CSI}}$, the entropy decays from $\ln 2$ to $S_{\mathrm{P}}$ of the spin-ice plateau ($0.01\lesssim\frac{T}{J}\lesssim0.1$) and $\chi_\parallel$ steeply decays to zero. This signal a crossover from a high-$T$ local-moment regime to a CC liquid or CSI regime. Here, the vanishing $\chi_\parallel$ indicates that QSI monopoles no longer survive in the QMC worldlines from $\tau=0$ to $1/T$. Finite-size effects and statistical errors are pronounced when this occurs at around $\frac{T}{J}\sim0.05$. Well below this temperature, a coherence of the gauge fields is expected to develop. Hence, we perform $\frac{1}{L^4}$-extrapolations~\cite{supplement}, as expected from emergent ``photons'' for QC liquids~\cite{hermele2004}, of the QMC data on $\varepsilon$ to $L\to\infty$ and then its numerical temperature derivative to obtain the specific heat, as plotted with black solid curves in the insets of Fig.~\ref{fig:2}(a) and (b), respectively.
On cooling below $T\sim0.01J$, the specific heat starts to increase again and exhibits another broad peak at $T_{\mathrm{QSI}}\sim0.001J$ where the Pauling entropy has been significantly released (Fig.~\ref{fig:2}(c)). The extrapolated data of $\varepsilon$ and $C$ for $T\le T_{\mathrm{QSI}}$ can be fitted in the asymptotic forms $\varepsilon(T=0)+\varepsilon_{\mathrm{ph}}(T)$ and $\frac{\partial \varepsilon_{\mathrm{ph}}}{\partial T}$, as shown with red dotted curves in the insets of Fig.~\ref{fig:2}(a) and (b), respectively, indicating that spin ice is melted by quantum fluctuations~\cite{onoda:09}. Here, $\varepsilon_{\mathrm{ph}}(T)=\frac{2a^3}{4(2\pi)^3}\int_0^{\frac{2\pi}{a}}d\bm{k}\frac{\xi_{\mathrm{ph}}(\bm{k})}{e^{\xi_{\mathrm{ph}}(\bm{k})/T}-1}$ is the energy density of photons in the noncompact pure U(1) gauge theory for Eq.~\eqref{eq:model} with the cubic lattice constant $a$, the energy dispersion $\xi_{\mathrm{ph}}(\bm{k})=\frac{2\hbar c}{a}\sqrt{\sum_{\mu<\nu}\sin^2\bm{k}\cdot\frac{\bm{b}_\mu-\bm{b}_\nu}{2}}$~\cite{benton2012} at the wavevector $\bm{k}$, and the ``light'' velocity $c$.
This implies that on cooling across $T_{\mathrm{QSI}}$, the system crosses over to a QC liquid~\cite{hermele2004} or QSI regime. 
Extremely careful experiments taking into account a long thermal relaxation time on high-quality samples might uncover this QSI regime with the entropy significantly reduced from $S_{\mathrm{P}}$, as recently tackled for Pr$_2$Zr$_2$O$_7$~\cite{kimura:13}, Pr$_2$Ir$_2$O$_7$~\cite{tokiwa:14}, and Dy$_2$Ti$_2$O$_7$~\cite{pomaranski:13}.
The above fitting yields $c\simeq 1.49(4) \frac{ag}{\hbar}$ with $g \equiv |\frac{3 J_\perp^3}{2 J^2}|$ for $\frac{J_\perp}{J} = -\frac{1}{11}$. This and the previous estimate $c = 1.8(1) \frac{ag}{\hbar}$ for $\frac{J_\perp}{J} = -\frac{1}{9.7}$~\cite{banerjee2008} show an enhancement by a factor of 2-3 from $c = 0.6(1) \frac{ag}{\hbar}$ in the asymptotic limit $|\frac{J_\perp}{J}|\to0$~\cite{benton2012}, indicating the importance of higher-order corrections in $\frac{J_\perp}{J}$. Since $T_{\mathrm{QSI}}$ should be governed by the energy scale of ``photons'', we conjecture $T_{\mathrm{QSI}}\simeq\frac{2\hbar c}{3a}$ to reproduce our results for $\frac{J_\perp}{J}=-\frac{1}{11}$.
 The QSI regime below $T_{\mathrm{QSI}}$ is actually adiabatically connected to the CSI regime, and hence it is not a deconfined phase in a strict sense, on the contrary to the mean-field result~\cite{savary:13}. It is most likely that the deconfinement occurs only at $T=0$. 
Then, the QC liquid ground state is never an exclusive quantum-mechanical superposition of spin ice rule states, as evidenced by a small but finite value of $\langle n^2\rangle$ and its upturn below $T\sim0.01J$ (Fig.~\ref{fig:2}(d)). This upturn reflects that states outside the spin-ice manifold are required for gaining the kinetic energy of spinons. Even in the pure U(1) gauge theory associated with Eq.~(1) ~\cite{hermele2004}, the perturbed wave function should acquire a finite QSI monopole density through a unitary transformation.


Now we clarify how spin correlations evolve on cooling in the above case of $\frac{J_\perp}{J}=-\frac{1}{11}$. Figure~\ref{fig:3}(a) presents the energy-integrated $\bm{Z}$-polarized neutron-scattering cross-sections on the $\bm{k}=\frac{2\pi}{a}(h,h,l)$ plane for non-Kramers cases like Pr, Ho, and Tb moments,
$
\sigma_{\rm SF}(\bm{k}) \equiv  \sigma_{\rm T}(\bm{k}) - \sigma_{\rm NSF}(\bm{k})
$ in the spin-flip (SF) channel, 
$
\sigma_{\rm NSF}(\bm{k}) \equiv \sum_{\mu,{\mu'}} \langle s^z_{\mu\; \bm{k}} s^z_{{\mu'}\; -\bm{k}} \rangle
[(\hat{\bm{k}}\times\hat{{\bm b}}_\mu)\cdot(\hat{\bm{k}}\times\bm{Z})] 
[(\hat{\bm{k}}\times\hat{{\bm b}}_\nu)\cdot(\hat{\bm{k}}\times\bm{Z})] 
$ in the non-spin-flip (NSF) channel, and the total
$
\sigma_{\rm Tot}(\bm{k}) \equiv \sum_{\mu,{\mu'}} \langle s^z_{\mu\; \bm{k}} s^z_{{\mu'}\; -\bm{k}} \rangle
[
\hat{{\bm b}}_\mu \cdot \hat{{\bm b}}_{\mu'} - (\hat{{\bm b}}_\mu \cdot \hat{\bm{k}})  (\hat{{\bm b}}_{\mu'} \cdot \hat{\bm{k}})
]$
without the nuclear form factor, where
$s^z_{\mu \bm{k}} \equiv \frac{1}{L^{3/2}}\sum_{{\bm R}_+} s^z_{{\bm R}_+ +{\bm b}_\mu /2} e^{i \bm{k}\cdot ( {\bm R}_+ + {\bm b}_\mu /2 )}$,
with ${\bm Z} \equiv \frac{1}{\sqrt{2}}\left(1,-1,0 \right)$, $\hat{{\bm b}}_\mu \equiv \frac{{\bm b}_\mu}{|{\bm b}_\mu|}$, and $\hat{\bm{k}}\equiv\frac{\bm{k}}{k}$. 
At $T>T_{\mathrm{CSI}}$, a broad SF scattering intensity appears along [100] and [111], as demonstrated for $\frac{T}{J}=1$ in Fig.~\ref{fig:3}(a). This elucidates experimental observations in Pr$_2$Zr$_2$O$_7$~\cite{kimura:13}.
On cooling below $T_{\mathrm{CSI}}\sim0.2J$ down to $0.1J$, the pinch-point singularity~\cite{isakov:04,henley:05} develops only in $\sigma_{\mathrm{SF}}$ at every reciprocal lattice vectors but $\bm{k}=(0,0,0)$, as shown for $\frac{T}{J}=0.1$ in Fig.~\ref{fig:3}(a). $\sigma_{\mathrm{SF}}$ around (002) becomes anisotropic on cooling from $\frac{T}{J}=4$ (red) to $0.05$ (green), as clearly seen by comparing Figs.~\ref{fig:3}(b) and (c). 
The pinch points never evolve into a real singularity, because the 2-in, 2-out spin ice rule is dynamically violated by the spin-flip exchange processes of Eq.~\eqref{eq:model}~\cite{onoda:09}. The $\bm{k}$ dependence of $\sigma_{\mathrm{NSF}}$ is invisible above $T\sim0.05J$, as is also clear from Figs.~\ref{fig:3}(b) and (d) as well as Fig.~\ref{fig:3}(a). On further cooling below $\frac{T}{J}=0.01$ where the entropy starts being reduced from $S_{\mathrm{P}}$ (Fig.~\ref{fig:2}(c)), the intensity of both $\sigma_{\mathrm{SF}}$ and $\sigma_{\mathrm{NSF}}$ around the reciprocal lattice vectors starts decaying, as shown in the two lower panels of Fig.~\ref{fig:3}(a). At $\frac{T}{J}=0.005$ and 0.002, they nearly follow the $\bm{k}$,$T$-dependent form of the photon contribution in the associated noncompact pure U(1) gauge theory~\cite{benton2012}, which is drawn with blue and violet dashed curves, respectively, in Figs.~\ref{fig:3}(b), (c), and (d). Nevertheless, we clearly observe the following deviations from the photon form. (i) On cooling, anisotropic peaks of $\sigma_{\mathrm{SF}}$ at reciprocal lattice vectors, e.g., (002) and (222), sharpen on top of the decaying photon contribution, while the height is saturated. (ii) Obviously, the disappearance of the pinch-point feature and the emergence of the broad [111]- and [100]-rod scattering intensity at $T\ge T_{\mathrm{CSI}}$ can never be described by the photon form. They all should be ascribed to effects of QSI monopoles.

\begin{figure}[htbp]
  \includegraphics[trim = 0 10 0 10, clip,width=7.5cm]{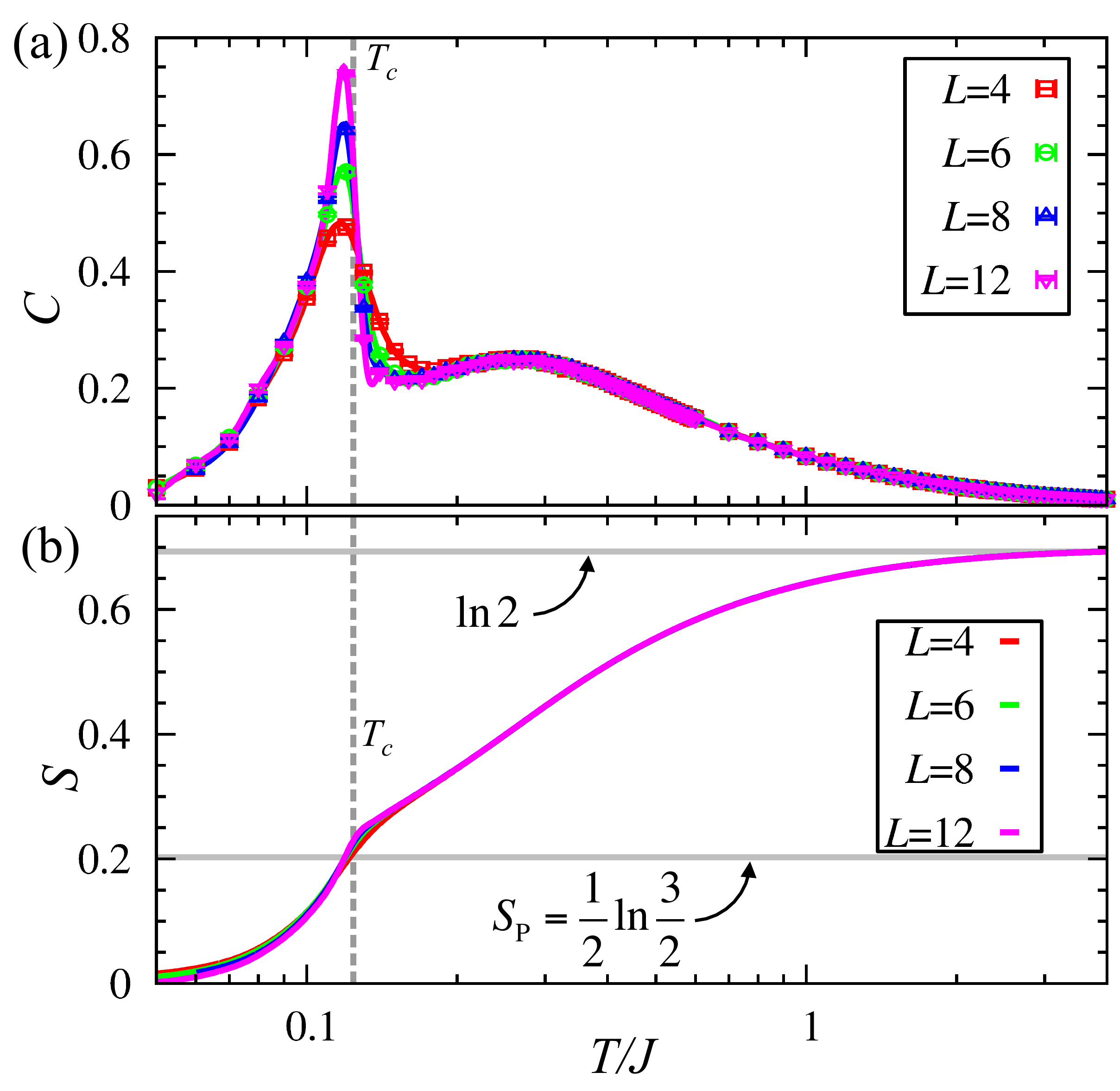} 
  \caption{
    (Color online)
    Temperature dependence of (a) $C$ and (b) $S$ for $\frac{J_\perp}{J}=-\frac{1}{5} < (\frac{J_\perp}{J})_c$. Dashed lines in (a) are the cubic spline interpolations of the QMC data. The entropy is computed from the cubic spline interpolation of the specific heat data with $\frac{T_{\rm max}}{J} = 4$. 
    \label{fig:4}
  }
\end{figure}

Now we focus on the case of $J_\perp<J_{\perp c}$, where a phase transition occurs to the XY-FM~\cite{banerjee2008}. 
We observe a clear discontinuous jump of the energy density $\varepsilon$ when $J_\perp$ is close to $J_{\perp c}$, 
while we observe only a sharp specific heat peak gradually growing with $L$ when $J_\perp$ is far from $J_{\perp c}$ (Fig.~\ref{fig:4}(a)). 
Our results suggest a possibility of either a weakly first-order or a second-order transition in the case of $\frac{J_\perp}{J}\leq -\frac{1}{6}$ \cite{supplement}.
The high-$T$ crossover at $T_{\mathrm{CSI}}\sim0.2J$ evidenced by a broad specific heat peak is also observed in this case, as marked with (green) dashed line in Fig.~\ref{fig:1} and demonstrated for $\frac{J_\perp}{J}=-\frac{1}{5}$ in Fig.~\ref{fig:4}~(a). However, the entropy $S$ (Fig.~\ref{fig:4}~(b)) does not show the spin ice plateau, which is masked by a spiky peak in $C$ due to a ferromagnetic transition at $T_c=0.124(3)J$.

The neutron-scattering profile in this case of $\frac{J_\perp}{J}=-\frac{1}{5}$ has also been computed above $T_c$. The results look almost the same as shown in Fig.~\ref{fig:3}: remnants of the pinch-point singulartity survive at $T=0.2J$, as indeed observed in Yb$_2$Ti$_2$O$_7$ slightly above $T_c$ where a first-order transition occurs to a nearly collinear ferromagnet~\cite{chang:11}. In this regard, Yb$_2$Ti$_2$O$_7$ slightly above $T_c$ is possibly only at an onset to a narrow, if any, QSI regime, and hence it is unlikely to observe ``photons'' in Yb$_2$Ti$_2$O$_7$.

Note that recent experiments on Tb$_2$Ti$_2$O$_7$~\cite{Taniguchi2013} have revealed a phase diagram which looks compatible with our result, 
except that our transverse spin order is interpreted as a quadrupole order~\cite{onoda:10,lee2012}.

\begin{acknowledgments}
{\bf Acknowledgements}:
The authors thank M. Hermele for stimulating discussions.
Numerical calculations were conducted on the RIKEN Integrated Cluster of Clusters.
This work was partially supported by Grants-in-Aid for Scientific Research under Grant No. 26800199 and No. 24740253 and by the RIKEN iTHES project.
S.O. acknowledges the hospitality of the Aspen Center for Physics supported by the National Science Foundation under Grant No. PHYS-1066293, and the hospitality of Nordita, Nordic Institute of Theoretical Physics, during his stays at the end of the work.
\end{acknowledgments}

\bibliographystyle{apsrev}
\bibliography{qsiarxiv}


  \topmargin -2.25cm
  \textheight = 26cm
  \marginparwidth = 0cm
%
%
%

\pagebreak
\widetext
\begin{center}
\textbf{\large Supplemental Material for ``
    Numerical evidence of quantum melting of spin ice: quantum-classical crossover''}
\end{center}
\setcounter{equation}{0}
\setcounter{figure}{0}
\setcounter{table}{0}
\setcounter{page}{1}
\makeatletter
\renewcommand{\theequation}{S\arabic{equation}}
\renewcommand{\thefigure}{S\arabic{figure}}

\begin{center}
In this Supplemental Material, we present magnified views of Figs.~2 in the main text for a higher visibility, 
$\frac{1}{L^4}$-extrapolations of the quantum Monte-Carlo data on the energy density as displayed in the inset of Fig.~2(a),
and a detailed analysis for determining the ferromagnetic transition temperature in the phase diagram shown in Fig.~1.
All the detailed descriptions are in the caption of each figure.
\end{center}

\maketitle

\begin{figure}[htb]
  \includegraphics[trim = 0 0 0 0, clip,width=0.9\textwidth]{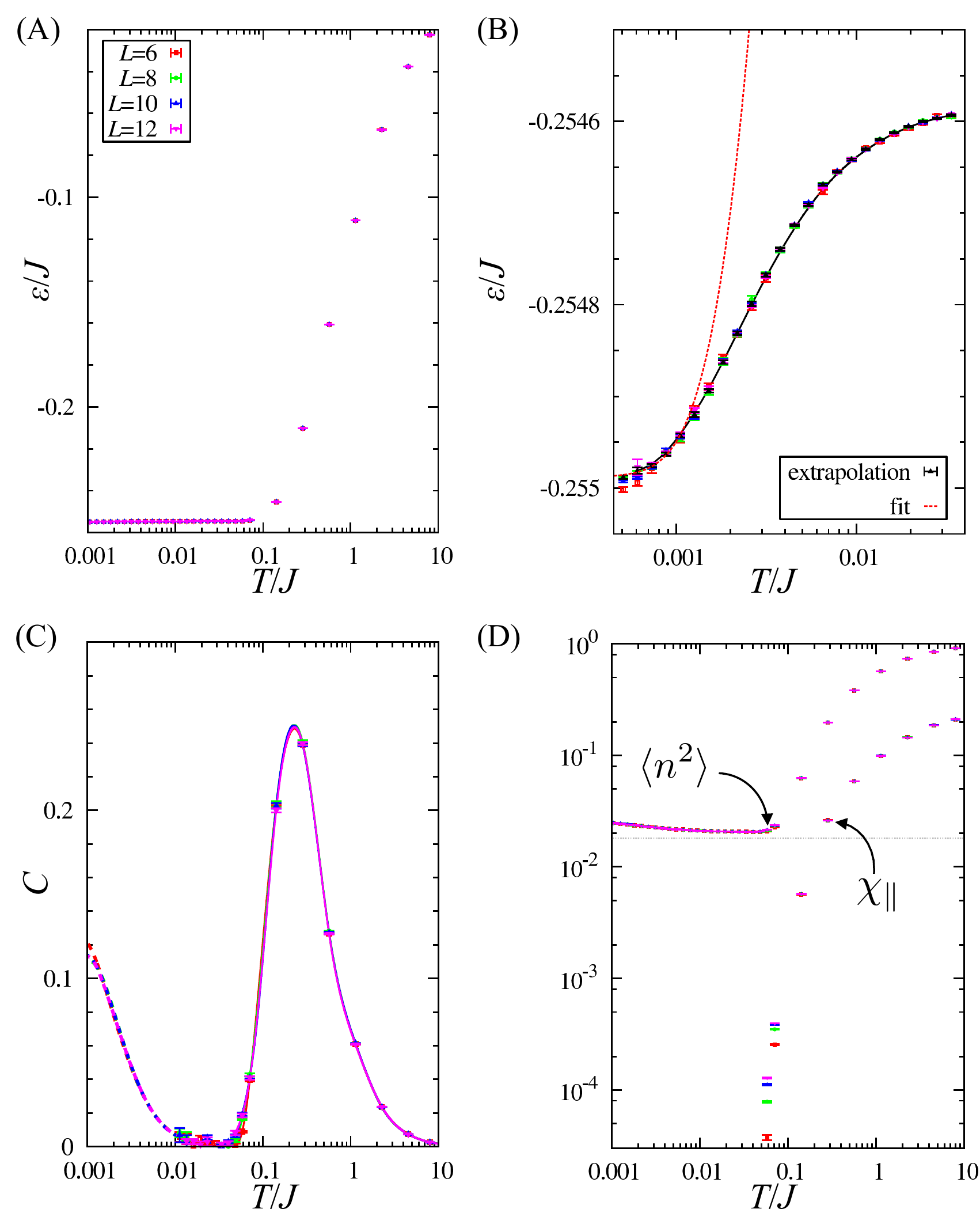} 
  \caption{
    Magnified views of (A) Fig.~2(a), (B) the inset of Fig.~2(a), (C) Fig.~2(b) and (D) Fig.~2(d) of the maintext.
    \label{fig:s1}
  }
\end{figure}

\begin{figure}[htb]
  \includegraphics[trim = 0 0 0 0, clip,width=\textwidth]{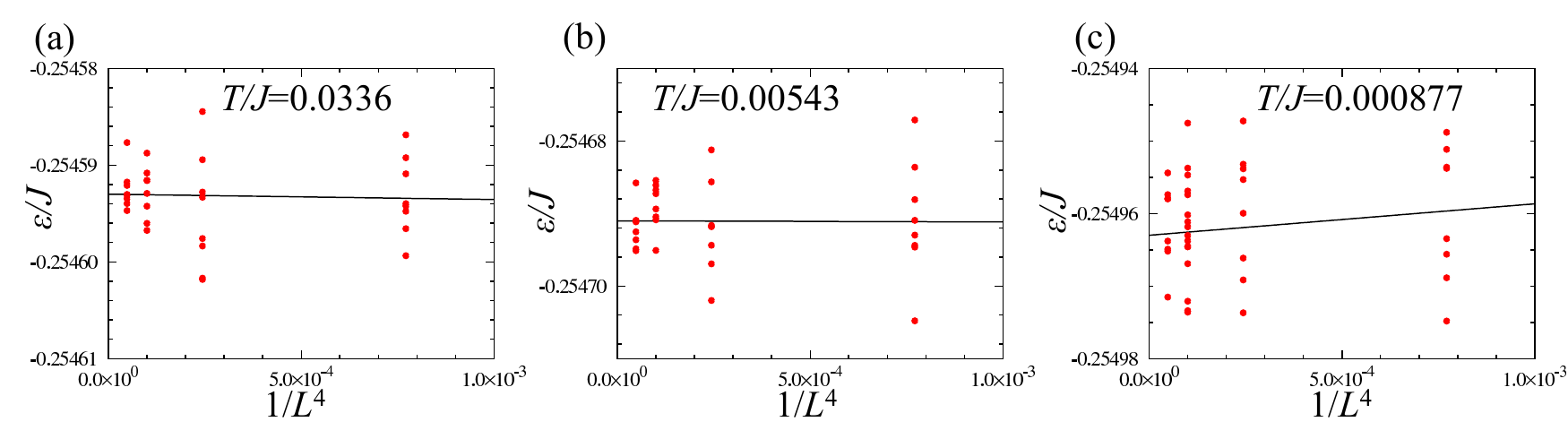} 
  \caption{
    Demonstration of the $\frac{1}{L^4}$-extrapolation of the energy density collected with quantum Monte-Carlo simulations for $L=6, 8, 10$, and 12, which are presented in the inset of Fig.~2(a), at
    (a) $T/J=0.0336$, 
    (b) $T/J=0.00543$, and
    (c) $T/J=0.000877$.
    Quantum Monte-Carlo data on the energy density has been collected into eight bins for each parameter set. Then, we perform a least-square fitting of the results into the form $\varepsilon_{\frac{1}{L}\to0}(T)+\tilde{\varepsilon}(T)\frac{1}{L^4}$ with a couple of adjustable parameters $\varepsilon_{\frac{1}{L}\to0}(T)$ and $\tilde{\varepsilon}(T)$ at all the temperature points below $0.05J$ for $\frac{J_\perp}{J}=-\frac{1}{11}$. The results of $\varepsilon_{\frac{1}{L}\to0}(T)$ are shown by black points with error bars in the inset of Fig.2(a) of the main text.
    \label{fig:s2}
  }
\end{figure}

\begin{figure}[htb]
  \includegraphics[trim = 0 0 0 0, clip,width=\textwidth]{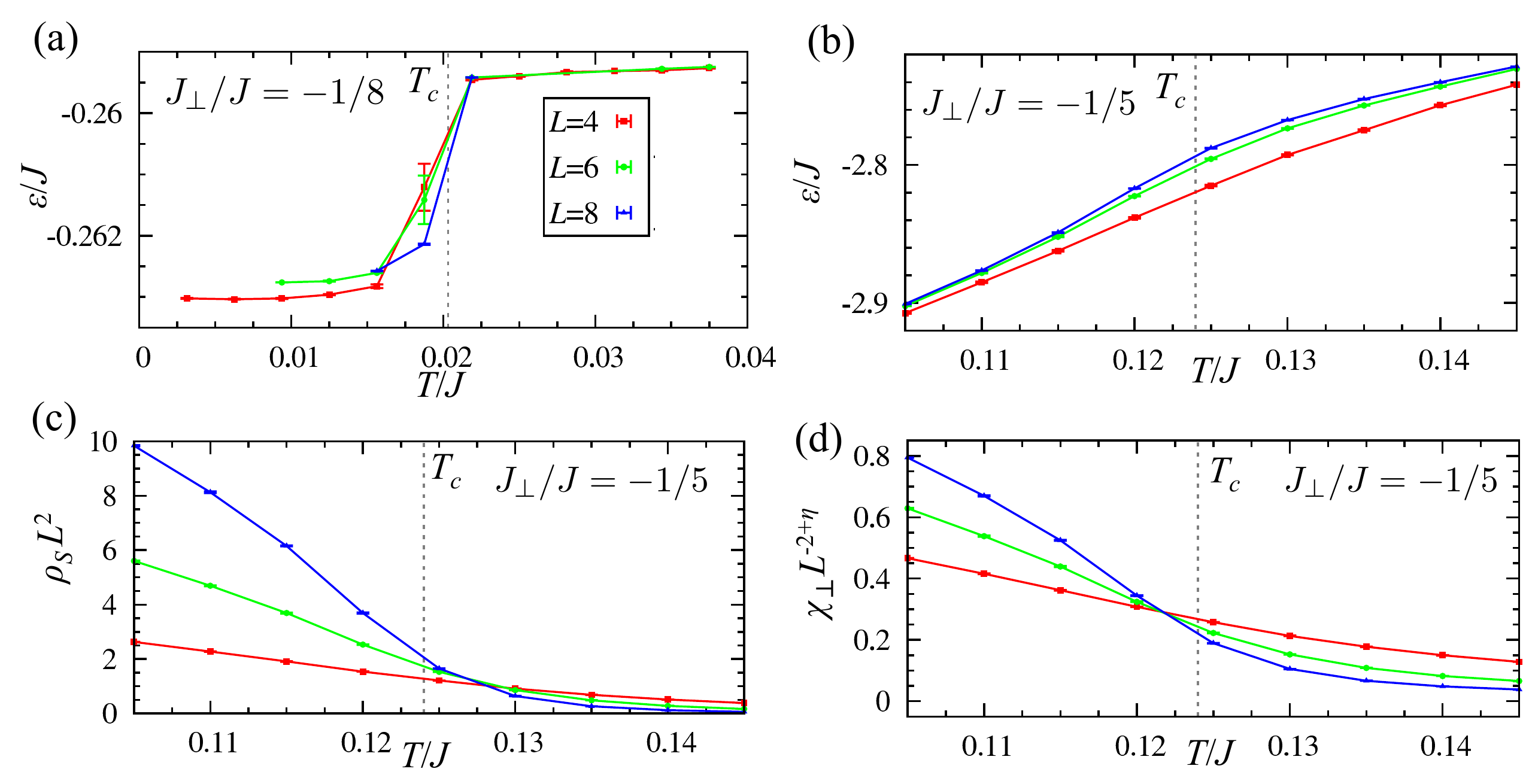} 
  \caption{
    Analysis for determining the ferromagnetic transition temperature.
    Each panel shows the temperature dependence of (a) the energy density $\varepsilon$ for $\frac{J_\perp}{J}=-\frac{1}{8}$,
    (b) $\varepsilon$,
    (c) the spin stiffness $\rho_S$ and 
    (d) the transverse spin susceptibility 
    $\chi_{\bot}
    \equiv\frac{1}{\beta} \int_0^\beta d\tau 
    \sum_{\bm r}
    \langle s^x_{\bm r}(\tau) s^x_{\bf 0}(0) \rangle$
    scaled by $L^{-2+\eta}$, where we assumed $\eta=0.038$ 
    for the 3D XY universality class~\cite{campostrini2001} for $\frac{J_\perp}{J}=-\frac{1}{5}$. 
    When $J_\perp$ is close to $J_{\perp c}$, $\varepsilon$ exhibits a clear discontinuous jump [the panel (a)], which gives a first-order transition temperature 
    $\frac{T_c}{J}=0.020(2)$.
    With negatively increasing $\frac{J_\perp}{J}$, $T_c$ increases, and the discontinuous jump associated with the first-order phase transition  becomes less clear [the panel (b)]. 
    Then, $T_c$ is estimated from the average of the two crossing temperatures of the scaled $\rho_S$ and $\chi_\bot$ [the panels (c) and (d)]. This value of $T_c$ reasonably coincides
    with the specific heat peak temperature. The difference in the two crossing temperature is taken as the error in the estimation of $T_c$ in Fig.~1 of the main text.
    Because of the slightly different crossing temperatures, a finite-size scaling does not hold for $L=4,6,8$ in this case.
    Our results suggest a possibility of either a weakly first-order or a second-order transition when $\frac{J_\perp}{J}\leq -\frac{1}{6}$. 
    \label{fig:s3}
  }
\end{figure}

\end{document}